\documentclass[twoside,twocolumn,9pt]{article}
\usepackage{extsizes}
\usepackage[super,sort&compress,comma]{natbib} 
\usepackage[version=3]{mhchem}
\usepackage[left=1.5cm, right=1.5cm, top=1.785cm, bottom=2.0cm]{geometry}
\usepackage{balance}
\usepackage{widetext}
\usepackage{times,mathptmx}
\usepackage{sectsty}
\usepackage{graphicx} 
\usepackage{lastpage}
\usepackage[format=plain,justification=raggedright,singlelinecheck=false,font={stretch=1.125,small,sf},labelfont=bf,labelsep=space]{caption}
\usepackage{float}
\usepackage{fancyhdr}
\usepackage{fnpos}
\usepackage[english]{babel}
\usepackage{array}
\usepackage{droidsans}
\usepackage{charter}
\usepackage[T1]{fontenc}
\usepackage[usenames,dvipsnames]{xcolor}
\usepackage{setspace}
\usepackage[compact]{titlesec}

\usepackage[markup=underlined]{changes}
\usepackage{todonotes}
\usepackage{multirow}
\usepackage{xcolor}
\usepackage{gensymb}

\definecolor{cream}{RGB}{222,217,201}

\begin{document}

\pagestyle{fancy}
\thispagestyle{plain}
\fancypagestyle{plain}{

\fancyhead[C]{\includegraphics[width=18.5cm]{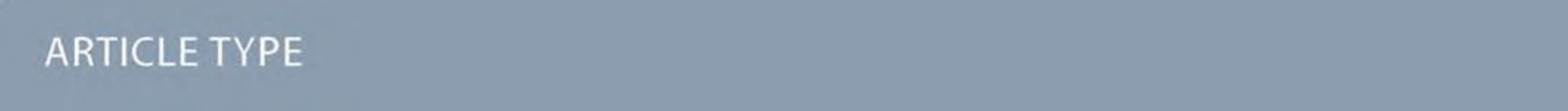}}
\fancyhead[L]{\hspace{0cm}\vspace{1.5cm}\includegraphics[height=30pt]{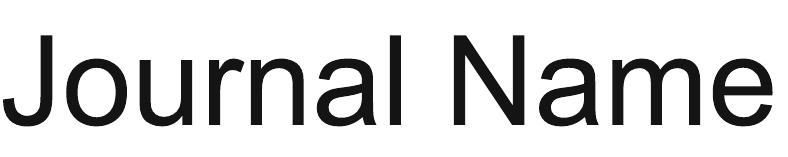}}
\fancyhead[R]{\hspace{0cm}\vspace{1.7cm}\includegraphics[height=55pt]{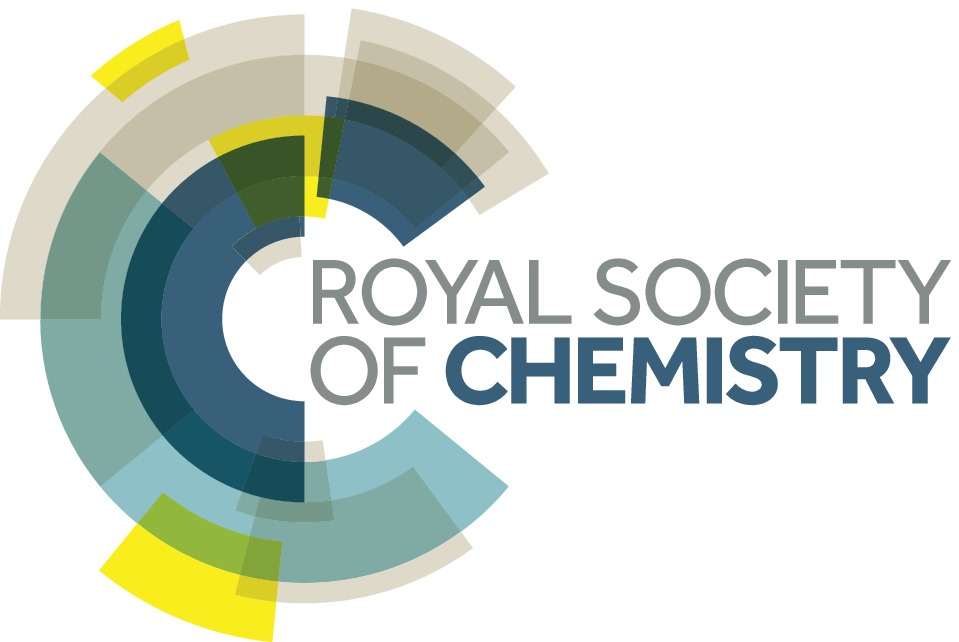}}
\renewcommand{\headrulewidth}{0pt}
}

\makeFNbottom
\makeatletter
\renewcommand\LARGE{\@setfontsize\LARGE{15pt}{17}}
\renewcommand\Large{\@setfontsize\Large{12pt}{14}}
\renewcommand\large{\@setfontsize\large{10pt}{12}}
\renewcommand\footnotesize{\@setfontsize\footnotesize{7pt}{10}}
\makeatother

\renewcommand{\thefootnote}{\fnsymbol{footnote}}
\renewcommand\footnoterule{\vspace*{1pt}%
\color{cream}\hrule width 3.5in height 0.4pt \color{black}\vspace*{5pt}} 
\setcounter{secnumdepth}{5}

\makeatletter 
\renewcommand\@biblabel[1]{#1}            
\renewcommand\@makefntext[1]%
{\noindent\makebox[0pt][r]{\@thefnmark\,}#1}
\makeatother 
\renewcommand{\figurename}{\small{Fig.}~}
\sectionfont{\sffamily\Large}
\subsectionfont{\normalsize}
\subsubsectionfont{\bf}
\setstretch{1.125} 
\setlength{\skip\footins}{0.8cm}
\setlength{\footnotesep}{0.25cm}
\setlength{\jot}{10pt}
\titlespacing*{\section}{0pt}{4pt}{4pt}
\titlespacing*{\subsection}{0pt}{15pt}{1pt}

\fancyfoot{}
\fancyfoot[LO,RE]{\vspace{-7.1pt}\includegraphics[height=9pt]{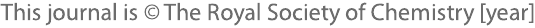}}
\fancyfoot[CO]{\vspace{-7.1pt}\hspace{13.2cm}\includegraphics{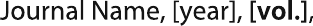}}
\fancyfoot[CE]{\vspace{-7.2pt}\hspace{-14.2cm}\includegraphics{head_foot/RF}}
\fancyfoot[RO]{\footnotesize{\sffamily{1--\pageref{LastPage} ~\textbar  \hspace{2pt}\thepage}}}
\fancyfoot[LE]{\footnotesize{\sffamily{\thepage~\textbar\hspace{3.45cm} 1--\pageref{LastPage}}}}
\fancyhead{}
\renewcommand{\headrulewidth}{0pt} 
\renewcommand{\footrulewidth}{0pt}
\setlength{\arrayrulewidth}{1pt}
\setlength{\columnsep}{6.5mm}
\setlength\bibsep{1pt}

\makeatletter 
\newlength{\figrulesep} 
\setlength{\figrulesep}{0.5\textfloatsep} 

\newcommand{\topfigrule}{\vspace*{-1pt}%
\noindent{\color{cream}\rule[-\figrulesep]{\columnwidth}{1.5pt}} }

\newcommand{\botfigrule}{\vspace*{-2pt}%
\noindent{\color{cream}\rule[\figrulesep]{\columnwidth}{1.5pt}} }

\newcommand{\dblfigrule}{\vspace*{-1pt}%
\noindent{\color{cream}\rule[-\figrulesep]{\textwidth}{1.5pt}} }

\makeatother

\twocolumn[
  \begin{@twocolumnfalse}
\vspace{3cm}
\sffamily
\begin{tabular}{m{4.5cm} p{13.5cm} }

\includegraphics{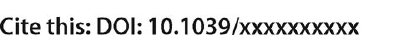} & \noindent\LARGE{\textbf{Towards Rechargeable Zinc-Air Batteries with Aqueous Chloride Electrolytes$^\dag$}} \\
\vspace{0.3cm} & \vspace{0.3cm} \\

 & \noindent\large{Simon Clark,\textit{$^{a,b,c}$} Aroa R. Mainar,\textit{$^{d}$} Elena Iruin,\textit{$^{d}$} Luis C. Colmenares,\textit{$^{d}$} J. Alberto Bl\'azquez,\textit{$^{d}$} Julian R. Tolchard,\textit{$^{c}$} Arnulf Latz,\textit{$^{a,b,e}$}} and Birger Horstmann$^{\ast}$\textit{$^{a,b,e}$} \\

\includegraphics{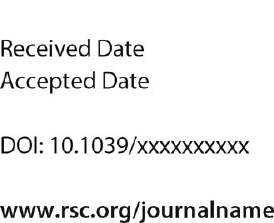} & \noindent\normalsize{This paper presents a combined theoretical and experimental investigation of aqueous near-neutral electrolytes based on chloride salts for rechargeable zinc-air batteries (ZABs). The resilience of near-neutral chloride electrolytes in air could extend ZAB lifetime, but theory-based simulations predict that such electrolytes are vulnerable to other challenges including pH instability and the unwanted precipitation of mixed zinc hydroxide chloride products. In this work, we combine theory-based simulations with experimental methods such as full cell cycling, operando pH measurements, ex-situ XRD, SEM, and EDS characterization to investigate the performance of ZABs with aqueous chloride electrolytes. The experimental characterization of near-neutral ZAB cells observes the predicted pH instability and confirms the composition of the final discharge products. Steps to promote greater pH stability and control the precipitation of discharge products are proposed.} \\

\end{tabular}

 \end{@twocolumnfalse} \vspace{0.6cm}

  ]

\renewcommand*\rmdefault{bch}\normalfont\upshape
\rmfamily
\section*{}
\vspace{-1cm}


\footnotetext{\textit{$^{a}$~German Aerospace Center (DLR), Pfaffenwaldring 38, 70569 Stuttgart, Germany. Fax: +49 (0)731 5034011 ; Tel: +49 (0)711 68628254; E-mail: birger.horstmann@dlr.de}}
\footnotetext{\textit{$^{b}$~Helmholtz Institute Ulm (HIU), Helmholtzstr. 11, 89081 Ulm, Germany. }}
\footnotetext{\textit{$^{c}$~SINTEF Industry, Richard Brikelands vei 2b, 7034 Trondheim, Norway. }}
\footnotetext{\textit{$^{d}$~CIDETEC Energy Storage,  P$^\circ$ Miram\'on, 196, Donostia-San Sebasti\'an 20014, Spain. }}
\footnotetext{\textit{$^{e}$~Ulm University (UUlm),  Albert-Einstein-Allee 47, 89081 Ulm, Germany. }}

\footnotetext{\dag~Electronic Supplementary Information (ESI) available: [details of any supplementary information available should be included here]. See DOI: 10.1039/b000000x/}



\section{Introduction}
Rechargeable zinc-air batteries (ZABs) are a promising post-Lithium-Ion battery technology\cite{Fu2017,Pang2017,Li2017b} for applications ranging from renewable energy storage, to electric vehicles\cite{Cano2018} and flexible electronics\cite{Tan2017b}.
Current state-of-the-art ZABs feature an alkaline electrolyte like \ce{KOH} for its high conductivity, good electrochemical reaction kinetics, and moderate \ce{Zn} solubility\cite{Mainar2018,Xu2015}. Unfortunately, the absorption of \ce{CO2} from air into the electrolyte leads to the parasitic formation of carbonates (\ce{CO3^{2-}}), which slowly poisons the electrolyte\cite{Stamm2017, Li2014}. For this reason, the lifetime of alkaline ZABs is cut short by a few weeks of continuous exposure to air. 

Engineering solutions to the carbonation challenge have been proposed\cite{Pei2014a}. The use of \ce{CO2} filters to scrub the feed-gas could delay the onset of carbonation\cite{Drillet2001}, but to reach competitive lifetimes, the \ce{CO2} concentration would need to be reduced by two orders-of-magnitude\cite{Stamm2017}. Mechanically rechargeable Zn-air fuel cells and \ce{Zn}-air flow batteries, in which the electrolyte is routinely replaced, have also been demonstrated\cite{Zhu2016a, Ma2015, Oh2018,Pei2014,Wang2014,Pichler2018}. These solutions are effective for some applications, but they add cost and complexity to the system. Ideally, the carbonation challenge should be addressed on the materials level.

Aqueous electrolytes with near-neutral pH values are resilient towards carbonation and could improve ZAB lifetime~\cite{An2018Heterostructure-PromotedElectrolyte}. The most common near-neutral electrolyte (NNE) is \ce{ZnCl2-NH4Cl}, which has been used in zinc-based LeClanch\'{e} batteries for over 100 years\cite{Heise1952,Garche2009}. In this tradition, we refer to zinc-air batteries with aqueous \ce{ZnCl2-NH4Cl} as LeClanch\'{e} zinc-air batteries (L-ZABs). 

The L-ZAB concept was first proposed in the 1970s\cite{Jindra1973}, but it has only recently become a broadly pursued topic in industry and research. Start-up companies are beginning to commercialize L-ZAB technology for grid-scale stationary applications\cite{Amendola2012}, and recent experimental research\cite{ThomasGoh2014,Sumboja2016} has verified the favorable cycling stability and lifetime of these systems. Although the future of L-ZABs is hopeful, there are some factors limiting their further development.

We recently performed a theoretical investigation of L-ZAB cell operation\cite{Clark2017}. Our continuum model confirms that the LeClanch\'{e} electrolyte is generally valid for ZAB applications and highlights some potential obstacles. The model predicts that the pH of the electrolyte can become acidic during charging and that mixed zinc salts, not \ce{ZnO}, generally dominate the discharge product. The instability of the pH exacerbates material degradation, and the precipitation of non-\ce{ZnO} products consumes the electrolyte and lowers the practical energy density of the cell. The study suggests that reducing the total chloride content and tuning the initial pH to be slightly alkaline (\emph{e.g.} pH 8) could improve the pH stability and support the precipitation of more favorable products. 

In this work, we combine experimental characterization with theory-based simulations to validate our understanding of L-ZABs. We build upon the previously reported thermodynamic analysis to show how LeClanch\'{e} electrolytes can be formulated and prepared to provide a stable pH value during operation and favor more desirable discharge products. Based on this analysis, we identify 4 electrolyte compositions for experimental investigation. Through the use of long-term cell cycling, operando electrolyte pH measurements, and ex-situ XRD, SEM, and EDS characterization of discharged and charged \ce{Zn} electrodes, we evaluate the effect of electrolyte composition on L-ZAB performance.

\section{Theory of LeClanch\'{e} Zinc-Air Batteries}

\begin{figure}[t!]
  \includegraphics[width=1.0\linewidth]{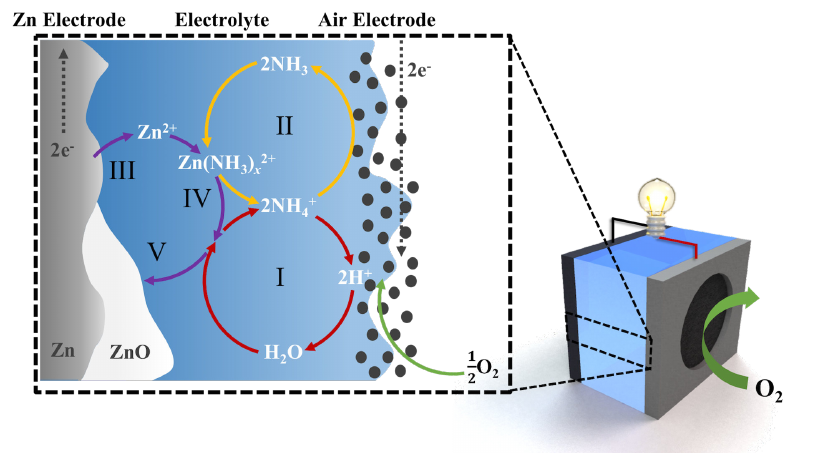}
  \caption{Schematic of idealized L-ZAB discharge. (I) The ORR occurs at the three-phase boundary of the air electrode. (II) The buffer reaction proceeds to stabilize the pH. (III) The zinc electrode dissolves to create \ce{Zn^{2+}}, which (IV) forms complexes with other solutes. (V) When the saturation limit of \ce{Zn^{2+}} is reached, zinc products precipitate. Possible solid discharge products include \ce{ZnO}, \ce{Zn(OH)2}, \ce{ZnCl2*4Zn(OH)2*H2O}, and \ce{ZnCl2*2NH3}.}
  \label{fgr:Schematic}
\end{figure}


In this section, we review the operating principle of L-ZABs and discuss how both the equilibrium and dynamic behavior of the electrolyte can influence cell performance. 

\subsection{Operating Principle}

Figure \ref{fgr:Schematic} shows the operational schematic of an idealized L-ZAB. During discharge, the oxygen reduction reaction (ORR) occurs at the so-called three-phase boundary of the porous bi-functional air electrode (BAE) with the help of a catalyst like \ce{MnO2}. The ORR drives a change in the concentration of \ce{H+} at the air electrode, and the pH is stabilized by the deprotonation of the weak acid \ce{NH4+}. The \ce{Zn} electrode is electrochemically oxidized to \ce{Zn^{2+}} ions, which form complexes with other solutes (\emph{e.g.} \ce{Cl-}, \ce{NH3}, or \ce{OH-}). The equations and standard redox potentials for the electrochemical reactions in the L-ZAB are
\begin{align}
&\ce{Zn} \rightleftharpoons \ce{Zn^{2+} + 2e-}, \ E^0 = -0.762 \: \textrm{V}, \\
&\ce{0.5O2 + 2H+ + 2e- <=> H2O}, \ E^0 = 1.229 \: \textrm{V}.
\end{align}
The stabilization of the electrolyte pH due to the weak acid buffer and the formation of zinc-amine complexes are described by the reactions,
\begin{align}
\ce{NH4+} &\ce{<=> NH3 + H+}, \: \textrm{and} \\
\ce{Zn^{2+} + \textit{x}NH3} & \ce{<=> \ce{Zn(NH3)_{\textit{x}}^{2+}}}.
\end{align}
When the solubility of \ce{Zn^{2+}} in the electrolyte is exceeded, zinc products precipitate. For the system to function as a true zinc-air battery\cite{Clark2017}, \ce{ZnO} should precipitate via
\begin{equation}
\ce{Zn^{2+} + H2O <=> ZnO(s) + 2H+},
\end{equation}
and give an overall reaction of
\begin{equation}
\ce{Zn + 0.5O2 <=> ZnO(s)}.
\end{equation}
However, as we will discuss in the following section, \ce{ZnO} is not always the dominant solid product. In some cases the discharge product can consist of a mix of \ce{ZnO}, \ce{Zn(OH)2}, \ce{ZnCl2*2NH3}, and \ce{ZnCl2*4Zn(OH)2*H2O}\cite{Zhang1996,Larcin1997,Passivation1976}. The overall cell reactions for various products are shown in Table \ref{tbl:OverallReactions}. The precipitation of non-\ce{ZnO} products consumes the electrolyte as an active material and reduces the energy density of the cell.

The performance of L-ZABs is governed by the delicate interplay between pH buffering, \ce{Zn^{2+}} chelation, and zinc salt precipitation. To better understand L-ZAB operation and identify optimum electrolyte formulations, we examine the thermodynamics of the system.

\begin{table}[t!]
  \caption{Overall cell reactions for different discharge products.}
  \label{tbl:OverallReactions}
  \def\arraystretch{1.5}
  \begin{tabular*}{\linewidth}{@{\extracolsep{\fill}}l}
    \hline
    Overall Reaction \\
    \hline
    \ce{Zn + 0.5O2 <=> ZnO(s)} \\
    \ce{Zn + 0.5O2  + H2O <=> Zn(OH)2(s)}  \\
    \ce{Zn + 0.5O2 + 2NH4+ + 2Cl- <=> ZnCl2*2NH3(s) + H2O} \\
    \ce{Zn + 0.5O2  + 0.8H2O + 0.4NH4+ + 0.4Cl- <=>} \\   \qquad \qquad \qquad \qquad\ce{0.2ZnCl2*4Zn(OH)2*H2O(s) + 0.4NH3}  \\
    \hline
  \end{tabular*}
\end{table}

\subsection{Equilibrium Thermodynamics}\label{sec:Thermodynamics}

In this section, we apply 0D thermodynamic models to investigate the equilibrium composition of the aqueous \ce{ZnCl2-NH4Cl-NH4OH} electrolyte, and discuss how this relates to L-ZAB operation. The models applied in this analysis are derived and validated in existing works\cite{Limpo1993,Limpo1995,Zhang2001,Vazquez-Arenas2012,Clark2017,Song2017} and described in the supplementary information$^\dag$. In the text, we use square brackets to denote concentration, \emph{e.g.} $[\ce{NH3}] = c_{\ce{NH3}}$ with units $\textrm{mol} \cdot \textrm{L}^{-1}$.

In aqueous solutions, the \ce{Zn^{2+}} ion forms complexes with other solutes\cite{Zhang1996}. The dominant zinc complex\cite{Zhang1996} in strongly alkaline electrolytes is the zincate ion, \ce{Zn(OH)4^{2-}}; the dominant zinc complex in acidic chloride electrolytes is the tetrachlorozincate ion, \ce{ZnCl4^{2-}}. But between the strongly acidic and strongly alkaline pH regions, the state of the zinc complex is very sensitive to changes in electrolyte composition.

\begin{figure}[t]
  \includegraphics[width=1.0\linewidth]{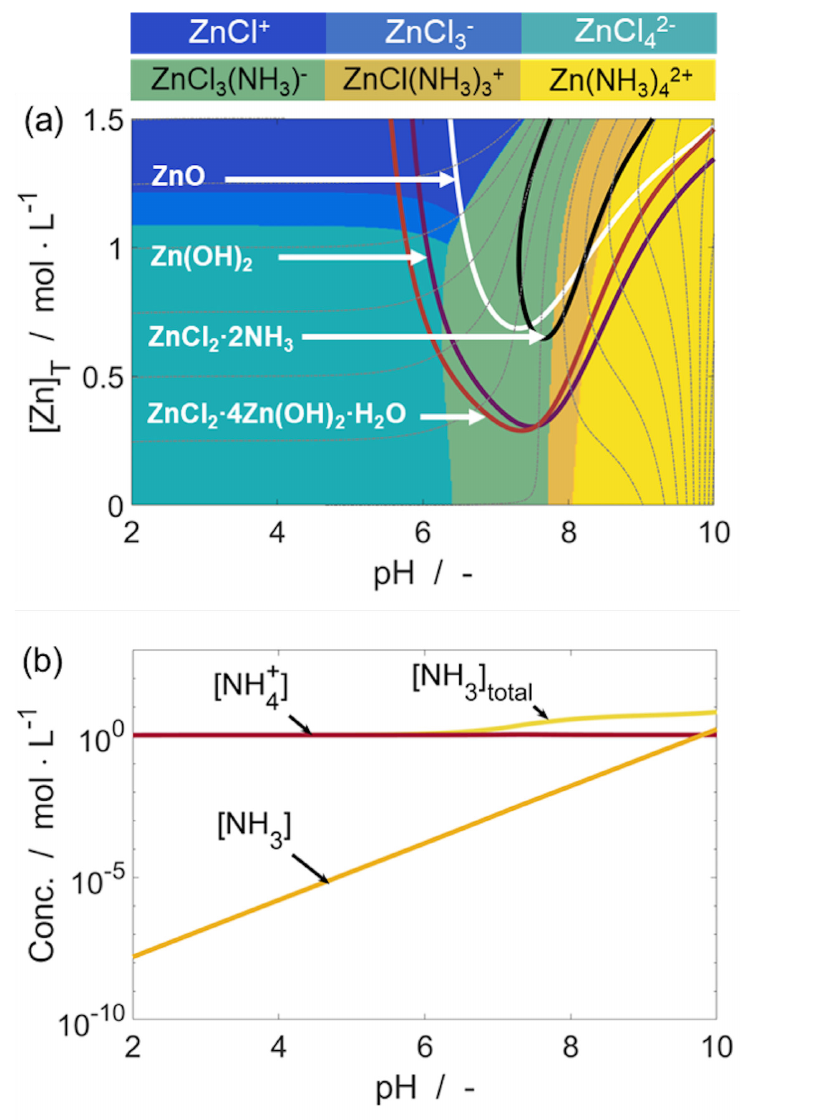}
  \caption{Speciation and solubility in the \ce{ZnCl2}-\ce{NH4Cl}-\ce{NH4OH} system. (a) Speciation of the \ce{Zn^{2+}} ion versus pH  and total zinc concentration (total chloride concentration, $[\ce{Cl}]_{\mathrm{T}}$, = 3 M). (b) \ce{NH3} distribution versus pH ($[\ce{Zn}]_{\mathrm{T}}$ = 1 M, $[\ce{Cl}]_{\mathrm{T}}$ = 3 M).}
  \label{fgr:Speciation}
\end{figure}

Figure \ref{fgr:Speciation}(a) shows the 2D zinc speciation and solubility landscapes for the \ce{ZnCl2-NH4Cl} system as a function of the pH and total concentration of zinc in solution, $[\ce{Zn}]_\textrm{T}$, for a fixed total chloride concentration, $[\ce{Cl}]_\textrm{T}$. The pH is adjusted through the addition of \ce{NH4OH}. The colored regions identify the dominant \ce{Zn^{2+}} complex, and the colored solid lines represent the solubility of various zinc products. The gray dashed lines represent fixed total \ce{NH3} concentrations (complexed \ce{NH3}, non-complexed \ce{NH3}, and \ce{NH4+}) and indicate paths the electrolyte follows as the ZAB is operated, as described by increases or decreases in the total zinc concentration. The concentrations of \ce{NH4+}, non-complexed \ce{NH3}, and total \ce{NH3} are shown in Figure \ref{fgr:Speciation}(b). 

In these diagrams, it is important to note the relationship between \ce{NH4+}, \ce{NH3}, and the zinc-amine complexes. We start by examining how the dominant zinc complex shifts as a function of pH, as shown in Figure \ref{fgr:Speciation}(a). For acidic pH values, the solution is dominated by zinc-chloride complexes because the concentration of \ce{NH3} is very low. The \ce{NH3} concentration rises with increasing pH values, leading to the formation of ternary zinc-chloride-amine complexes. When the concentrations of \ce{NH3} and \ce{NH4+} approach the equivalence point at pH 9.8, the solution is already dominated by \ce{Zn(NH3)4^{2+}}.

The formation of \ce{Zn(NH3)_{\textit{x}}^{2+}} complexes also has an important effect on the solubility of zinc products. \ce{ZnO} and \ce{Zn(OH)2} are normally insoluble in the near-neutral pH regime. However, \ce{NH3} is able to act as a chelator for \ce{Zn^{2+}} ions, thereby increasing the solubility of zinc products. 

Figure \ref{fgr:Speciation}(a) shows that in the acidic pH regime, \ce{Zn^{2+}} is very soluble. As the pH approaches the near-neutral regime, the solubility falls sharply until the increasing \ce{NH3} concentration becomes high enough to chelate the \ce{Zn^{2+}} ions. The solubility levels off and subsequently increases as the solution becomes saturated with \ce{NH3}. 

Figure \ref{fgr:Speciation}(a) can also be used to predict how electrolyte composition affects the stable working point of the battery (i.e. where the anodic, cathodic, and precipitation reactions form a complete cycle). During discharge, an electrolyte with an initial total zinc concentration of 0.5 M and pH of 4 will follow the gray dashed line until the solubility limit of \ce{ZnCl2*4Zn(OH)2*H2O} is reached around pH 6, and the battery achieves a stable working point. On the other hand, an electrolyte with the same initial total zinc concentration but with an initial pH of 9 will reach its stable working point around pH 8 and \ce{Zn(OH)2} is the first solid to precipitate. 

\begin{figure}[b!]
  \includegraphics[width=1.0\linewidth]{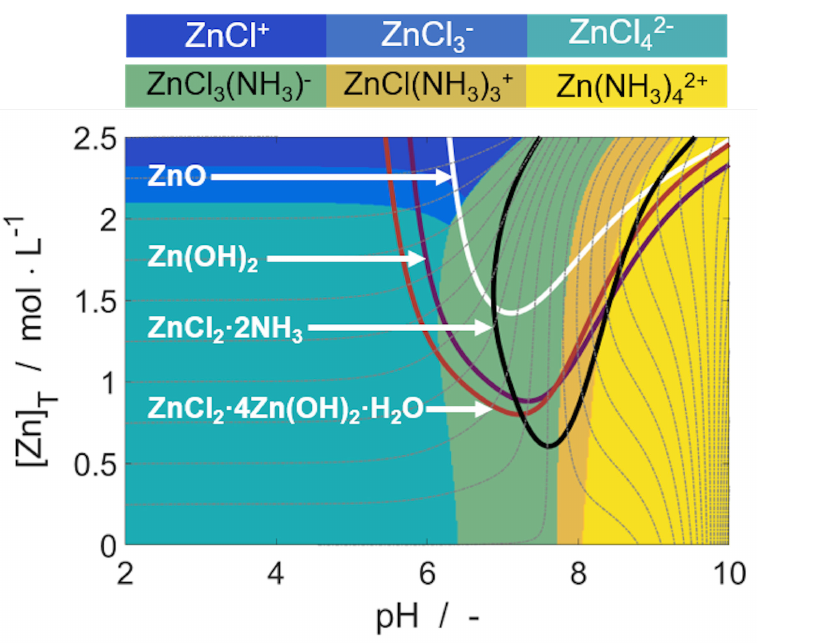}
  \caption{Speciation and solubility in the \ce{ZnCl2}-\ce{NH4Cl}-\ce{NH4OH} system, $[\ce{Cl}]_{\mathrm{T}}$ = 5 M.}
  \label{fgr:Thermo2D}
\end{figure}

The solubility of zinc products is also strongly linked to the concentration of chloride in the electrolyte. Increasing the total chloride content of the electrolyte, as shown in Figure \ref{fgr:Thermo2D}, decreases the solubility of chloride-rich products like \ce{ZnCl2*2NH3}. This is important for L-ZAB operation because it shows how changes in local electrolyte concentration and pH can alter the composition of the discharge product.

During L-ZAB operation, the pH of the electrolyte is stabilized by the buffer reaction \ce{NH4+ <=> NH3 + H+} and can be calculated in terms of the buffering species concentrations as 
\begin{equation}
\mathrm{pH} = \mathrm{pK_a}-\mathrm{log}_{10}\frac{[\ce{NH4+}]}{[\ce{NH3}]}.
\end{equation}
The capacity and reversibility of the buffer are described by the ratio $[\ce{NH4+}]:[\ce{NH3}]$. 

First we consider the capacity of the pH buffer. To achieve a stable operational pH, the value of the ratio $[\ce{NH4+}]:[\ce{NH3}]$ should be kept as constant as possible,
\begin{equation}
\frac{\partial\frac{[\ce{NH4+}]}{[\ce{NH3}]}}{\partial t} \approx 0, \qquad \frac{\partial\frac{[\ce{NH4+}]}{[\ce{NH3}]}}{\partial x} \approx 0.
\end{equation}
The buffer reaction consumes \ce{NH4+} and produces \ce{NH3} as it proceeds, causing the value of $[\ce{NH4+}]:[\ce{NH3}]$ to fall and creating a slow and steady increase in pH. Fortunately, the formation of complexes between \ce{NH3} and \ce{Zn^{2+}} allows the buffer reaction to proceed while the concentration of free \ce{NH3} remains relatively constant. In this way, the time-rate-of-change of $[\ce{NH4+}]:[\ce{NH3}]$ is reduced and the capacity of the buffer to stabilize the pH is enhanced. 

The practical reversibility of the buffer is described by the magnitude of $[\ce{NH4+}]:[\ce{NH3}]$. For most compositions in the near-neutral pH range, the concentration of \ce{NH4+} is much higher than \ce{NH3} (see Figure \ref{fgr:Speciation}(b)). This allows \ce{NH4+} to act as a proton donor and effectively buffer pH shifts in the alkaline direction. But if the pH becomes more acidic, there is only a small amount of \ce{NH3} available to act as proton acceptors. Although some excess \ce{NH3} can be supplied from \ce{Zn(NH3)_{\textit{x}}^{2+}} complexes, the reaction is very susceptible to any concentration gradients that could develop. Therefore, the buffer reaction can manage pH shifts in the alkaline direction, but the practical reversibility of the reaction to manage similar shifts in the acidic direction is limited. Because of this, there is a risk that the electrolyte could become acidic when the L-ZAB is charged. 

\begin{figure}[t!]
  \includegraphics[width=\linewidth]{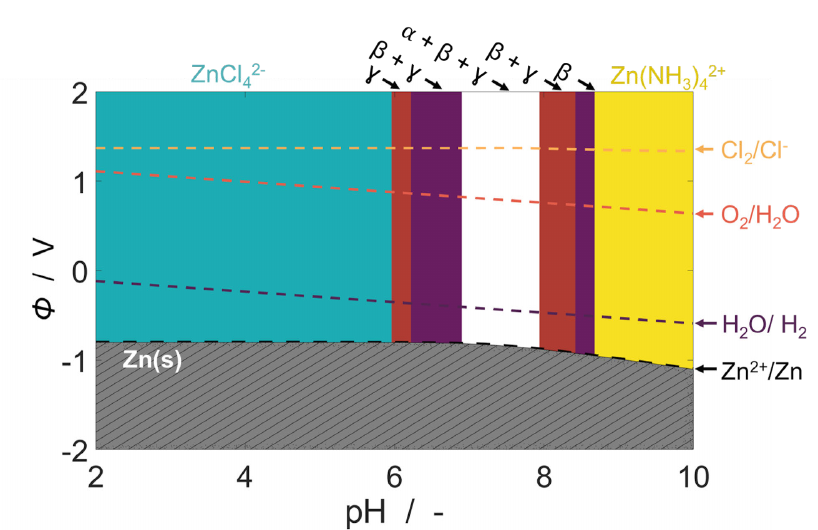}
  \caption{Calculated Pourbaix diagram for the aqueous \ce{ZnCl2-NH4Cl-NH4OH} system. $\alpha = \ce{ZnO}$, $\beta = \ce{Zn(OH)2}$, $\gamma = \ce{ZnCl2*4Zn(OH)2*H2O}$ ($[\ce{Cl}]_{\mathrm{T}}$ = 2.6M, $[\ce{Zn}]_{\mathrm{T}}$ = 0.5M).}
  \label{fgr:Pourbaix}
\end{figure}
 
Finally, we examine the equilibrium potentials of the electrochemical reactions. Figure \ref{fgr:Pourbaix} shows a Pourbaix diagram for the L-ZAB system. The equilibrium redox potential of the \ce{Zn} electrochemical reaction is below the redox potential for \ce{H2} evolution, indicating that the \ce{Zn} electrode is thermodynamically unstable in water. The kinetics of the hydrogen evolution reaction (HER) are slow and can be further suppressed by the addition of dopants like \ce{Hg}, \ce{In}, or \ce{Bi} to the \ce{Zn} electrode\cite{Lysgaard2018a,SureshKannan1995,Vorkapic1974,Baugh1983}. The equilibrium redox potential for chlorine gas evolution is close to the oxygen redox potential at very acidic pH values, but the two potentials separate as the pH becomes more alkaline. This supports the experimental observation from Sumboja, et al.\cite{Sumboja2016} that no \ce{Cl2} gas is evolved during charging. Zn passivation behavior occurs between pH values of circa 6-9, where mixed zinc products become insoluble. 

This analysis demonstrates how the stable working point of the battery can be predicted from thermodynamic considerations. Furthermore, it is shown that for most electrolyte compositions in the near-neutral pH regime, mixed zinc-chloride-hydroxide salts are most likely to precipitate. A three dimensional analysis of zinc speciation and solubility as a function of pH, \ce{NH4Cl}, and \ce{ZnCl2} concentrations is available in the supplementary information$^\dag$.

The equilibrium properties of the system are predicted from thermodynamics, but the real performance of L-ZAB cells deviate due to kinetics and mass transport limitations. In the following, we apply a dynamic model to consider these effects.

\subsection{Electrolyte Transport Dynamics}

\begin{figure}[b!]
  \includegraphics[width=1.0\linewidth]{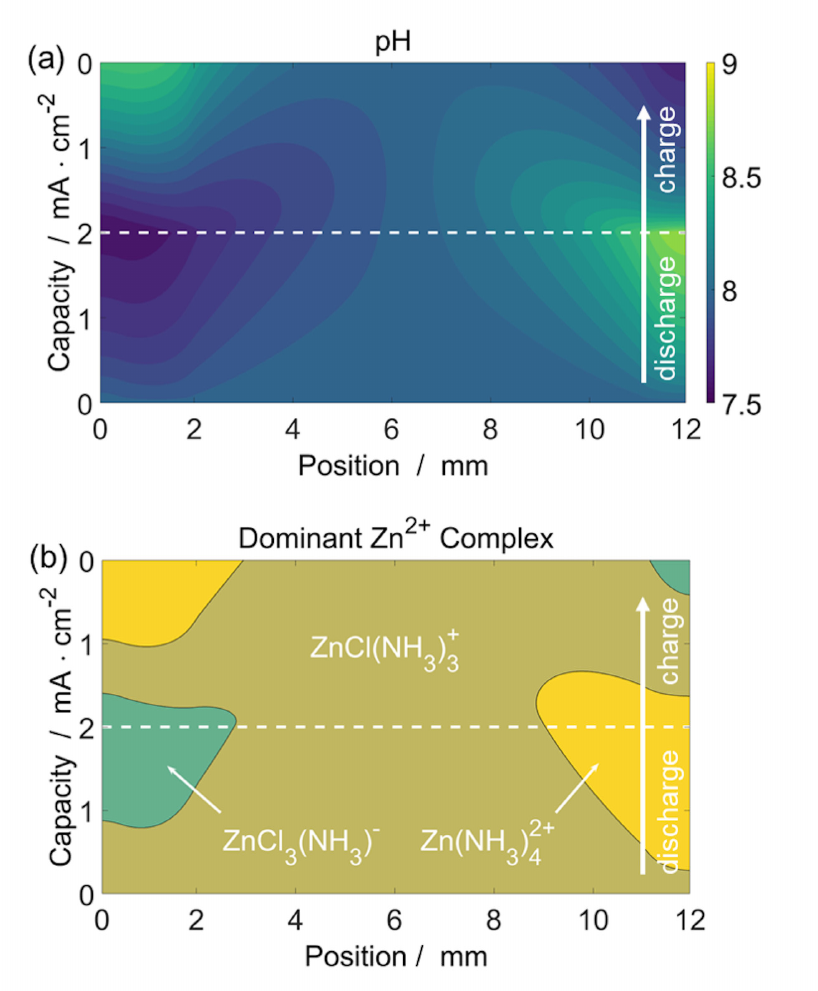}
  \caption{Dynamic profiles of (a) electrolyte pH and (b) dominant \ce{Zn^{2+}} complex in a L-ZAB cell over one discharge-charge cycle. The electrolyte is \ce{0.5MZnCl2-1.6MNH4Cl} pH 8. The cell is operated with current density $\mathrm{j_d} = \mathrm{j_c} = 1 \mathrm{mA}\cdot\mathrm{cm}^{-2}$.}
  \label{fgr:ZAB_Dyanmics}
\end{figure}

\begin{table*}[t!]
  \caption{Measured physicochemical properties of the four proposed electrolyte compositions compared with literature values for the standard alkaline ZAB electrolyte, 30 wt\% \ce{KOH}. Values for ionic conductivity (IC), mass density ($\rho$), dissolved oxygen concentration ([\ce{O2}]), and viscosity ($\mu$) are measured for each electrolyte.}
  \label{tbl:ElectrolyteProperties}
  \def\arraystretch{1.5}
  \begin{tabular*}{\textwidth}{@{\extracolsep{\fill}}ccccccccc}
    \hline
    Designation  & [\ce{ZnCl2}] & [\ce{NH4Cl}] & pH & IC ($\mathrm{mS} \cdot \mathrm{cm}^{-1}$) & $\rho$ ($\mathrm{g} \cdot \mathrm{mL}^{-1}$) & [\ce{O2}] ($\mathrm{mg} \cdot \mathrm{L}^{-1}$) & $\mu$ (cP) \\
    \hline
    E4   & 0.51 & 2.34 & 4 & 215 & 1.06 & 6.16 & 1.12 \\
    E6 & 0.51 & 2.34 & 6 & 206 & 1.08 & 7.12 & 1.09 \\
    E7 & 0.26 & 5.00 & 7 & 382 & 1.05 & 6.74 & 1.09 \\
    E8 & 0.50 & 1.60 & 8 & 209 & 1.05 & 6.61 & 1.15 \\
    \hline
    30 wt\% \ce{KOH} & - & - & 14.8 & 638\cite{Mainar2018a} & 1.28\cite{Akerlof1941} & 2.52\cite{Davis1967} & 2.23\cite{Sipos2000} \\
    \hline
  \end{tabular*}
\end{table*}

To simulate the dynamic performance of L-ZABs, we implement a 1D continuum model of the system and examine the performance over a discharge-charge cycle.

In the \ce{ZnCl2-NH4Cl-NH4OH} electrolyte, the large quantity of solute species combined with the orders-of-magnitude concentration swings that occur create difficulties for numerical solvers of traditional continuum models. Similar systems like the electrochemical desalination of water~\cite{Dykstra2017TheoryDeionization} or ammonia recovery from liquid bio-waste~\cite{Dykstra2014TheorySystems} have been successfully modeled, but there is a dearth of continuum models for L-ZAB performance. In our first model-based investigation of L-ZAB performance\cite{Clark2017}, we derived a novel framework for modeling transport in LeClanch\'{e} electrolytes. The framework defines a set of so-called quasi-particles that describe the quantities of mass and charge that are conserved in the homogeneous electrolyte reactions. In that way, the computational effort to obtain a solution is significantly improved. In an upcoming publication, we expand the validity of the quasi-particle framework to cover a wider range of electrolyte compositions beyond \ce{ZnCl2-NH4Cl-NH4OH}.

The quasi-particle model solves the equations for mass and charge continuity in the electrolyte. The concentration of each quasi-particle, $q$, is determined from the mass continuity equation, while the local electro-neutrality condition is set by the charge continuity equation. 

\begin{align}
\frac{\partial (c_q\varepsilon_{\mathrm{e}})}{\partial t} &= \underbrace{-\vec{\nabla}\cdot \vec{N}_q^{\mathrm{D,M}}-\vec{\nabla}\cdot \vec{N_q^{\mathrm{C}}}}_\text{transport} + \overbrace{\dot{s}_q}^\text{source}, \\
0 &= \underbrace{-\vec{\nabla}\cdot \vec{j}}_\text{transport} + \overbrace{\sum_iz_i\dot{s}_i}^\text{source}.
\end{align}
Detailed derivation, parameterization, validation, and discussion of the continuum modeling method is available in existing works \cite{Neidhardt2012, Horstmann2013, Stamm2017, Clark2017, Clark2018} and in the supplementary information$^\dag$ of this article~\cite{Newman,Shock1988,Shock1989,Shock1997,Frank1996a,Atkins2006,Livermore1990,Sverjensky1997,Limpo1993,Limpo1995,Clever1992,Reichle1975}.

Figure \ref{fgr:ZAB_Dyanmics} presents the performance of an L-ZAB cell with pH 8 0.5M \ce{ZnCl2} - 1.6M \ce{NH4Cl} over one discharge-charge cycle. The dynamic pH profile, shown in Figure \ref{fgr:ZAB_Dyanmics}(a), indicates that the pH in the BAE trends alkaline during discharging. On the other hand, when the cell is charged, the pH in the BAE trends acidic. In both cases, the buffer reaction is able to stabilize the pH in the near-neutral regime. At the \ce{Zn} electrode, the pH trends acidic during discharging. This is because the excess \ce{Zn^{2+}} takes up what small amount of \ce{NH3} is present. When the cell is charged, \ce{Zn^{2+}} is redeposited and releases \ce{NH3} into the electrolyte, causing the pH to trend alkaline. 

Figure \ref{fgr:ZAB_Dyanmics}(b) shows the dominant zinc complex in the solution. Initially the dominant complex over the entire cell domain is \ce{ZnCl(NH3)3^{+}}, but as the pH begins to shift and concentration gradients build up in the cell, the dominant complex becomes \ce{Zn(NH3)4^{2+}} in the air electrode and \ce{ZnCl3(NH3)-} in the \ce{Zn} electrode. Comparing the dynamic \ce{Zn} speciation with the equilibrium values calculated in the previous section shows how cell operation can affect the inhomogeneous behavior of the electrolyte.

The simulation results described in this section provide a foundation for understanding and interpreting experimental measurements of L-ZAB cells. In the following sections, we experimentally characterize L-ZABs with a variety of electrolyte compositions and compare the results with model-based predictions. 


\section{Experimental Methods}

\begin{table*}[t]
  \caption{Obtained electrochemical results during the cell cycling tests and corresponding electrolyte evaporation under open-circuit conditions. Overpotential is defined as $\Delta V = V_{\mathrm{OER}} - V_{\mathrm{ORR}}$.}
  \label{tbl:CyclingEvaporation}
  \def\arraystretch{1.1}
  \begin{tabular*}{\textwidth}{@{\extracolsep{\fill}}cccccc}
    \hline
    \textbf{Cycle Number} & & \textbf{E4} & \textbf{E6} & \textbf{E7} & \textbf{E8} \\
    \hline
    \multirow{4}{*}{1} & ORR (V) & 0.887 & 0.929 & 0.860 & 0.888 \\
    	& OER (V) & 2.079 & 2.067 & 1.977 & 1.922 \\
        & Overpotential (V) & 1.192 & 1.138 & 1.117 & 1.034 \\
        & Evaporation (wt\%) & 0.00 & 0.00 & 0.00 & 0.00 \\
    \hline
    \multirow{4}{*}{25} & ORR (V) & 0.924 & 0.909 & 0.915 & 0.947 \\
    	& OER (V) & 2.070 & 2.095 & 2.098 & 2.014 \\
        & Overpotential (V) & 1.146 & 1.186 & 1.183 & 1.067 \\
        & Evaporation (wt\%) & 1.70 & 0.57 & 0.00 & 5.87 \\
    \hline
    \multirow{4}{*}{50} & ORR (V) & 0.882 & 0.896 & 0.896 & 0.909 \\
    	& OER (V) & 2.069 & 2.088 & 2.117 & 2.047 \\
        & Overpotential (V) & 1.187 & 1.192 & 1.221 & 1.138 \\
        & Evaporation (wt\%) & 3.80 & 0.57 & 0.00 & 8.19 \\
    \hline
    \multirow{4}{*}{100} & ORR (V) & 0.848 & 0.847 & 0.866 & 0.761 \\
    	& OER (V) & 2.077 & 2.194 & 2.117 & 2.254 \\
        & Overpotential (V) & 1.229 & 1.347 & 1.251 & 1.493 \\
        & Evaporation (wt\%) & 5.34 & 1.69 & 0.00 & 12.58 \\
    \hline
    \multirow{4}{*}{150} & ORR (V) & 0.858 & 0.863 & 0.853 & 0.718 \\
    	& OER (V) & 2.079 & 2.172 & 2.142 & 2.272 \\
        & Overpotential (V) & 1.221 & 1.309 & 1.289 & 1.554 \\
        & Evaporation (wt\%) & 6.25 & 1.69 & 0.00 & 14.04 \\
    \hline
  \end{tabular*}
\end{table*}

\subsection{Preparation of materials}

Four different electrolyte systems (designated E4, E6, E7, and E8) were prepared from \ce{ZnCl2} (EMD Millipore 98 \%) and \ce{NH4Cl} (EMD Millipore 99.5 \%) dissolved in deionized water and the pH value was adjusted with \ce{NH4OH} (Fluka analytical 5.0 N). Table 2 lists the formulation of each electrolyte system and corresponding pH value. The electrolyte compositions were chosen to reflect existing studies in the literature~\cite{ThomasGoh2014,Sumboja2016} and to evaluate the recently proposed composition from Clark et al.~\cite{Clark2017}. Furthermore, the variation in selected electrolyte compositions are defined as to demonstrated changes in both pH stability and zinc precipitation product, as discussed in the section on Equilibrium Thermodynamics.


The bifunctional air electrode was prepared by mixing 70 wt.\% carbon nanotube (CNT, Arkema Graphistrength\textsuperscript{TM} C100), 20 wt.\% electrolytic manganese dioxide (EMD, Tosoh Hellas A. I. C.) and 10 wt.\% PTFE (Dyneon TF 5032 PTFE). The mixture was pressed twice for 1 minute at 50 bar against a carbon gas diffusion layer (Freudenberg H23C9). Once the electrodes were pressed, they were heated at 340$^\circ$C for 30 minutes where 2.2 mg cm$^{-2}$ of catalyst loading were achieved.

\subsection{\textbf{Physicochemical characterization of electrolytes}}

Physicochemical properties of the different electrolyte systems were analyzed with specific equipment for those measurements. In this context, the ionic conductivity (IC), viscosity ($\mu$), dissolved oxygen ([\ce{O2}]) and mass density ($\rho$) were measured for each formulation. The obtained values are listed in Table \ref{tbl:ElectrolyteProperties}.

\subsection{\textbf{Electrochemical characterization}}

In this work two customized electrochemical cell designs were used: ex-situ (C-EI) and operando pH (C-OpH). In the first cell (C-EI), the electrodes were separated by 0.9 cm and 1.1 mL of electrolyte was injected. The C-OpH cell requires more space between both electrodes to place two pH microelectrodes (Mettler Toledo, InLab$^{\text{\textregistered}}$ Micro) near the positive and negative electrodes. In this context, the C-OpH design features a distance of 2.8 cm between the electrodes and 4.4 mL of electrolyte. Photos of the cells are available in the supplementary information$^\dag$. In both electrochemical cells the bifunctional air electrode and a zinc foil (Alfa Aesar, 99.98\%, 250 $\mu$m thickness) were used as working and counter electrode, respectively, with an active area of 1.327 cm$^{2}$. The electrochemical analyses were carried out in a BaSyTEC Battery Test System.

Operando pH measurements were performed in the C-OpH cell design, applying a current density of 2 $\mathrm{mA} \cdot \mathrm{cm}^{-2}$ for 29 h of discharge and 29 h of charge (330 $\mathrm{mAh} \cdot \mathrm{g}^{-1}$; 40\% depth of discharge). 

Evaporation of the electrolyte was analyzed by measuring the weight loss over time at open-circuit conditions in the C-EI cell. The same cell design was used for the cycling tests of different electrolyte systems. In this case, a current density of 1 $\mathrm{mA} \cdot \mathrm{cm}^{-2}$ was applied during 2 hours discharge and 2 hours charge.

In order to evaluate the nature of the solid reaction products during cycling, two sets of experiments were performed. In the first, cells were discharged at a rate of 1 $\mathrm{mA} \cdot \mathrm{cm}^{-2}$ to a DoD of 40\%, and in the second cells were discharged in the same manner before being charged at 1 $\mathrm{mA} \cdot \mathrm{cm}^{-2}$ back to the original fully charged state.

\subsection {Physical Characterization of cell reaction products}

X-ray diffraction data were collected using a Bruker D8 Advance A25 powder diffractometer equipped with a Cu K-$\alpha$ radiation source and LynxEye XE\textsuperscript{TM} detector. All measurements were collected in Bragg-Brentano mode. Measurement of powder products after discharging to 40\% DoD was performed with the product still attached to the Zn foil anode, whilst powder products after discharging + charging were removed from the anode foil and dispersed on a "zero background" single crystal Si sample holder using silicon grease as adhesive. The phases present within the samples were identified using the ICDD PDF4+ 2017 crystal structure database~\cite{ICDDDatabase} and Crystallographic Open Database (COD)~\cite{Grazulis2009,Grazulis2012,Grazulis2015,LeBail2005,Downs2003}.  Phases were confirmed via Rietveld-type fitting, but it was not possible to adequately correct the complex preferred orientation exhibited by product phases and so final fitting was performed via Pawley-type whole powder pattern fitting in which only peak positions are constrained. All fitting was performed using the Bruker Topas v5 analysis software. 

Electron microscopy and EDS element mapping were performed using a Hitachi S3400N electron microscope equipped with an Oxford Instruments Aztec EDS system. Top surface images were collected from dry samples following electrochemical testing. For cross-sectional imaging, dried as-reacted anodes were embedded in epoxy resin (Struers EpoFix), and manually polished using grinding papers down to 4000 grit / 5$\mu$m grit size. In order to avoid dissolution or reaction, the samples were dry-polished without water or lubricant. Prior to imaging, both top and cross-section samples were coated with a thin layer of carbon to aid conductivity.

\section{Results and discussion}
\begin{figure}[t]
  \includegraphics[width=\linewidth]{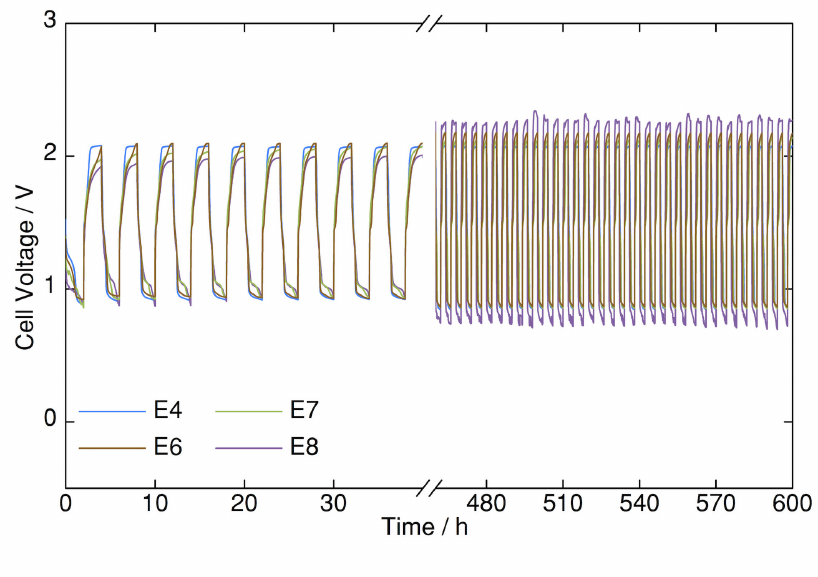}
  \caption{Cell voltage for L-ZABs cycled at current density $\mathrm{j_d} = \mathrm{j_c} = 1 \mathrm{mA}\cdot\mathrm{cm}^{-2}$ for $\mathrm{t_d} = \mathrm{t_c} = 2$ hours.}
  \label{fgr:ExpCycles}
\end{figure}

\subsection{Material Properties}

An overview of the physicochemical characterization of each electrolyte formulation is given in Table \ref{tbl:ElectrolyteProperties}. To provide an adequate frame of reference for these values, we compare them with properties of the standard electrolyte for alkaline ZABs (30 wt\% KOH) as reported in the literature~\cite{Mainar2018a,Akerlof1941,Davis1967,Sipos2000}. The ionic conductivity (IC) measurements indicate that electrolytes E4, E6, and E8 have comparable ionic conductivity values just above 200 $\mathrm{mS} \cdot \mathrm{cm}^{-1}$, while E7 shows a substantially higher conductivity of 382 $\mathrm{mS} \cdot \mathrm{cm}^{-1}$. This is likely due to the higher concentration of \ce{NH4Cl} in E7 as compared to the other electrolytes. Although the measured IC values are lower than that of \ce{KOH} (638 $\mathrm{mS} \cdot \mathrm{cm}^{-1}$), they are still in a suitable range for battery electrolyte applications. Analysis of the dissolved oxygen content (DO) and viscosity ($\mu$) of the electrolytes also indicate their suitability for ZAB applications. The dissolved oxygen levels are over twice as high as those found in \ce{KOH}, and the viscosity is roughly half that of \ce{KOH}. Higher dissolved oxygen concentration is beneficial for the kinetics of the ORR, and the low viscosity helps achieve good transport and wetting behavior in the air electrode. On the other hand, lower electrolyte viscosity could increase the risk of flooding the air electrode (see supplementary information$^\dag$). Care should be taken to adjust the hydrophilic/hydrophobic properties of the BAE substrate accordingly~\cite{Danner2014,Danner2016}.

\subsection{Full Cell Cycling}

Figure \ref{fgr:ExpCycles} compares the voltages of ZAB cells with the various electrolytes after 150 discharge-charge cycles over 600 hours at 1$\mathrm{mA}\cdot \mathrm{ cm}^{-2}$ (2h discharge, 2h charge). Electrolyte E8 presents an overpotential ($\Delta V = V_{\mathrm{OER}} - V_{\mathrm{ORR}}$) lower than 1.15 V, and is the lowest value compared with E4, E6 and E7 during the first 50 cycles (see Table \ref{tbl:CyclingEvaporation}). However, after 200 hours of battery cycling, electrolyte E8 shows a significant degradation in both the magnitude and the stability of the cell potential, while electrolytes E4, E6, and E7 show relatively stable performance during the cycling period. This may be due to increased loss of electrolyte by evaporation observed in E8 (12.58 wt\% in the 100th cycle, Table \ref{tbl:CyclingEvaporation}). Evaporation values have been taken at open-circuit conditions; they might be undervalued during the cycling testing due to a possible competition between OER and HER. The electrolyte evaporation in the C-EI cell design reduces the electrolyte level in contact with active materials. This reduces the practical active area of the BAE and, as a consequence, the real applied current density is increased, leading to a higher overpotential.

\subsection{Operando pH Stability}

\begin{figure}[b!]
  \includegraphics[width=1\linewidth]{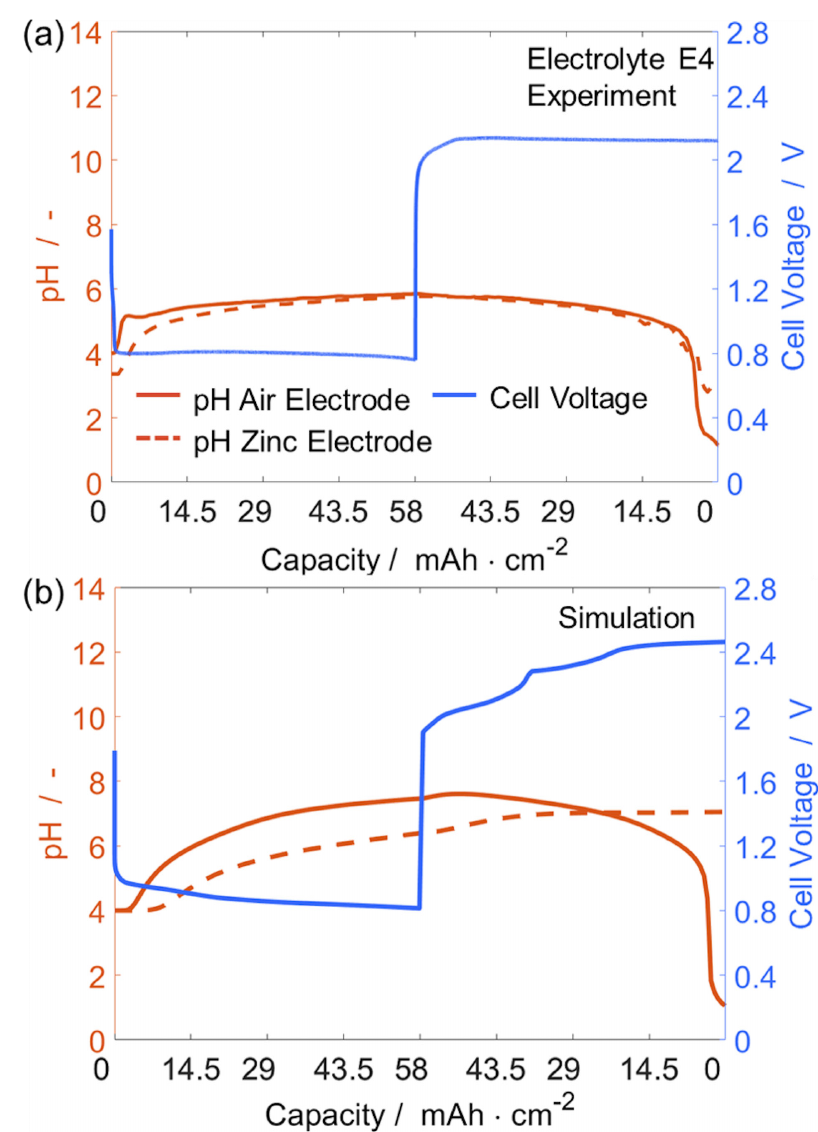}
  \caption{Comparison of measured and predicted pH profiles near the air electrode and the \ce{Zn} electrode from (a) experiment and (b) simulation. For a single cycle at current density $\mathrm{j_d} = \mathrm{j_c} = 2 \mathrm{mA}\cdot\mathrm{cm}^{-2}$.} 
  \label{fgr:Exp_pH}
\end{figure}

\begin{figure*}[t!]
  \includegraphics[width=1\textwidth]{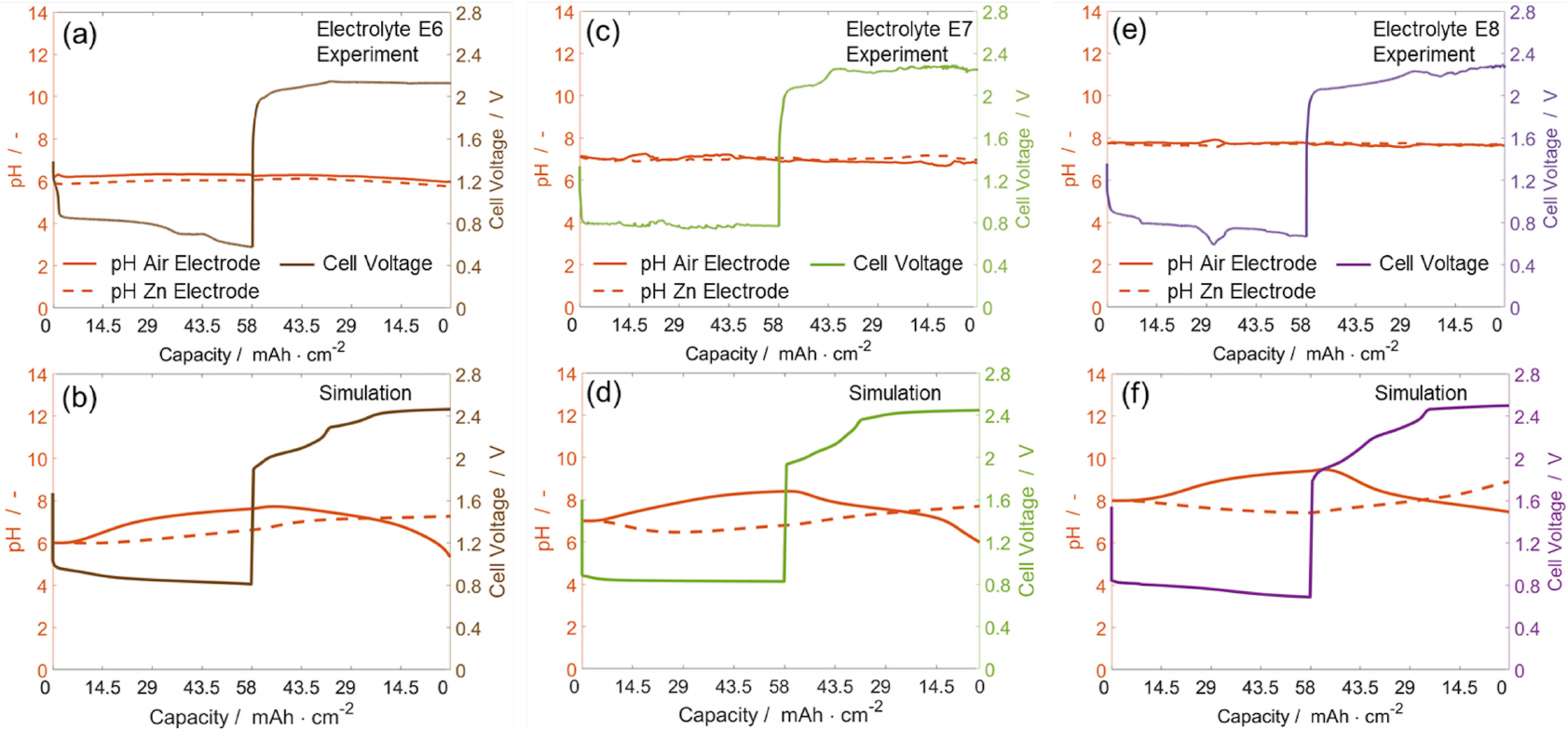}
  \caption{Comparison of measured and predicted pH profiles near the air electrode and the \ce{Zn} electrode from (a, c, e) experiment and (b, d, f) simulation. For a single cycle at current density $\mathrm{j_d} = \mathrm{j_c} = 2 \mathrm{mA}\cdot\mathrm{cm}^{-2}$.} 
  \label{fgr:Exp_pH2}
\end{figure*}

According to our understanding of L-ZAB performance as discussed in the theory section of this paper and our previous work\cite{Clark2017}, we predict that the electrolyte in the air electrode can become strongly acidic during charging due to the slow diffusion of \ce{NH3}. To validate this prediction, operando pH measurements are taken near the air and \ce{Zn} electrodes during a single discharge-charge cycle. Electrolyte E4 is the formulation most likely to become unstable because there is initially very little \ce{NH3} in the solution and the composition is far from the stable working point of the cell. Therefore, an L-ZAB featuring E4 offers the best opportunity to observe the predicted behavior.  

Figure \ref{fgr:Exp_pH}(a) shows the measured pH profiles in an L-ZAB with electrolyte E4. These curves contain two features of interest. First, although the electrolyte is initially at pH 4, there is a rapid increase at the start of discharge that eventually approaches a steady-state value near pH 6. The pH increase begins at the air electrode followed by a delayed increase at the \ce{Zn} electrode. Second, when the cell is charged, the pH is initially stable but drops to strongly acidic values near the end of charging. This drop begins in the air electrode and continues in the \ce{Zn} electrode. The pH in the \ce{Zn} electrode rebounds upward at the very end of charging. The L-ZAB model predicts this behavior~\cite{Clark2017}. The simulation results can help elucidate the mechanism behind the observed pH swings. 

Figure \ref{fgr:Exp_pH}(b) shows the simulated pH values in a L-ZAB with electrolyte E4. As observed in the experiment, the model predicts that the pH rapidly increases at the air electrode from the start of discharge until the cell reaches a stable working point in the near-neutral pH regime. A comparable shift is expected at the \ce{Zn} electrode, but it is delayed due to slow mass transport across the electrolyte bath and the excess concentration of \ce{Zn^{2+}}. As discharge continues, the rate of pH change stabilizes for both the BAE and the \ce{Zn} electrode. When the cell is charged, the pH near the BAE begins to drop and is stabilized by the buffer reaction. On the other side, the pH near the \ce{Zn} electrode becomes slightly more alkaline as \ce{Zn^{2+}} is deposited, releasing more \ce{NH3} from zinc-amine complexes. Near the end of charging, a \ce{NH3} mass transport limitation becomes dominant in the air electrode. With \ce{NH3} locally depleted, the buffer reaction is no longer effective and the pH drops to acidic values.

Measured and simulated pH curves for the remaining L-ZAB electrolyte systems are compared in Figure \ref{fgr:Exp_pH2}. In contrast to electrolyte E4, the pH of the other systems remains more stable. There is no increase at the start of discharge because electrolytes E6, E7, and E8 are formulated at their stable working points. The drop to acidic values at the end of charging is not observed at the measurement location in these systems.

There is generally good agreement between the predicted and observed pH behavior. The model tends to overestimate pH changes than are measured in the experiment. This may be because the measurement is taken at a single point in three-dimensional space, while the simulation is simplified to one-dimension. Because the pH is measured near the electrodes, there is a delay between the onset of pH variations in the electrodes and when they can be observed in the measurement. Nonetheless, the major predictions of the model including pH increase to the stable working point during discharging and the rapid fall to acidic values during charging are experimentally observed.

The pH fluctuations that occur during cell cycling can have important consequences for L-ZAB lifetime and performance. The suitability of ORR/OER catalyst materials is strongly dependent on electrolyte pH. For example, although \ce{MnO2} is a good catalyst in alkaline and neutral solutions, it is known to dissolve in acidic media~\cite{Huynh2014, Takashima2012a,Pokhrel2015}. Therefore, future research should examine ways to stabilize the pH in the air electrode during charging.

The electrolyte pH has a strong influence on the composition, precipitation, and dissolution of zinc salts. This, in turn, has a strong influence on the electrochemical performance of the \ce{Zn} electrode. In the following section, we examine the precipitation behavior of discharged and discharged-charged \ce{Zn} electrodes as a function of electrolyte composition.

\subsection{\ce{Zn} Electrode Characterization}

The dominant discharge product in a true \ce{Zn}-air cell should be \ce{ZnO}. The thermodynamic analysis shown in Figures \ref{fgr:Speciation} and \ref{fgr:Thermo2D} predicts that the precipitation product in \ce{ZnCl2-NH4Cl} electrolytes is a mix of \ce{ZnCl2*4Zn(OH)2*H2O}, \ce{ZnCl2*2NH3}, and \ce{Zn(OH)2}. To investigate this prediction, ex-situ XRD, SEM, and EDS measurements are performed on \ce{Zn} electrodes galvanostatically cycled between 0 - 40\% DoD in each electrolyte formulation. 

\begin{figure}[t]
  \includegraphics[width=1\linewidth]{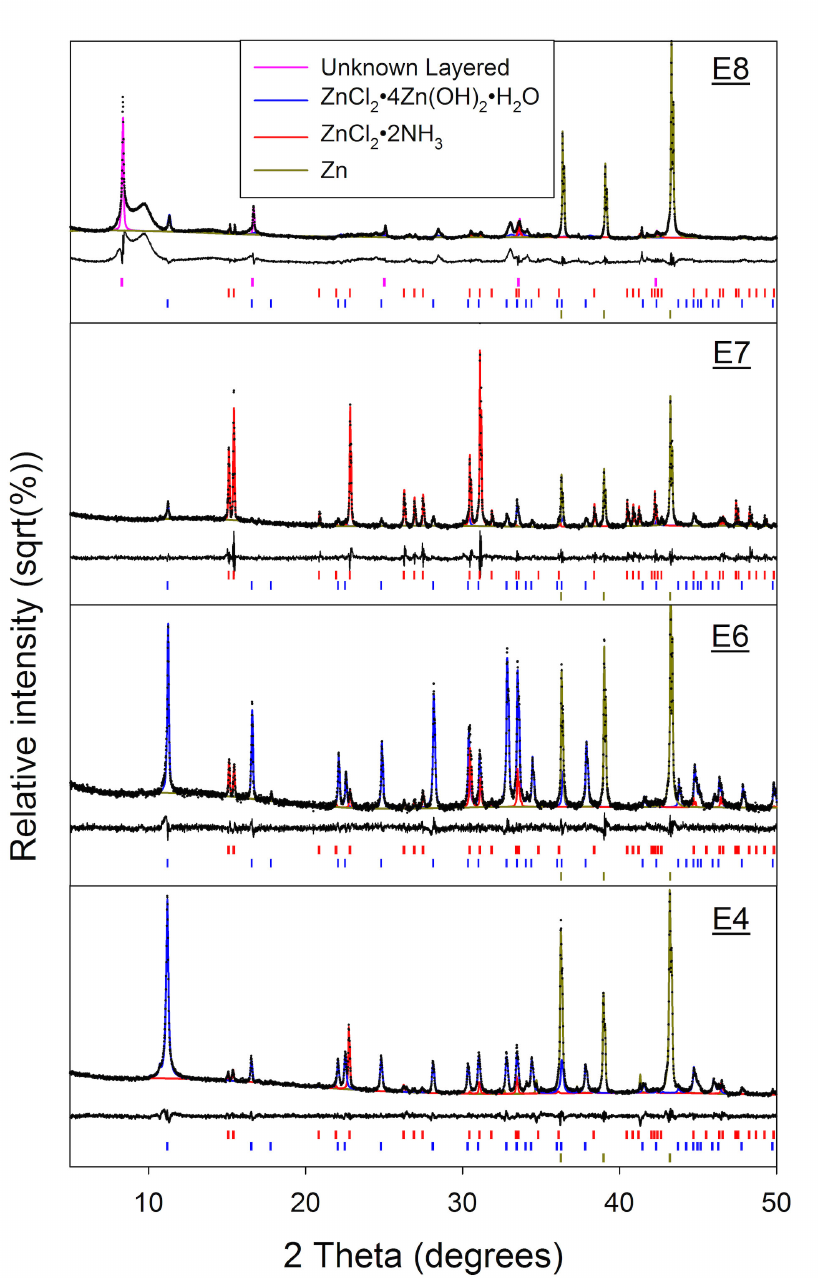}
  \caption{Powder X-ray diffractograms for anode product phases obtained at 40\% DoD in electrolytes E4, E6, E7, and E8. Data are scaled with the square-root of intensity to emphasize weak reflections. Samples were measured on the host Zn foil anode.} 
  \label{fgr:XRD}
\end{figure}

\begin{figure}[t]
  \includegraphics[width=1\linewidth]{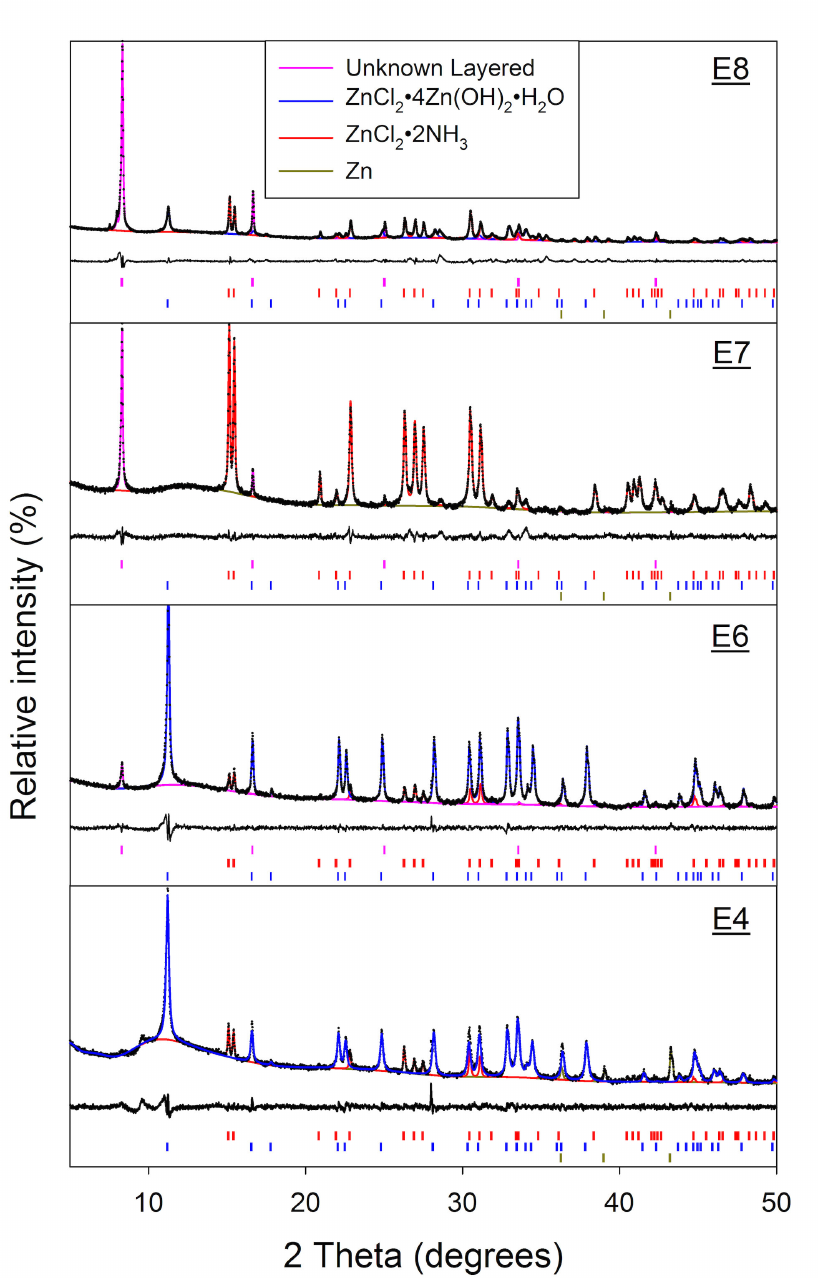}
  \caption{Powder X-ray diffractograms for anode product phases obtained from electrodes discharged 40\% DoD and charged in electrolytes E4, E6, E7, and E8. Data are scaled with the square-root of intensity to emphasize weak reflections.} 
  \label{fgr:XRD_Charged}
\end{figure}

Figures \ref{fgr:XRD} and \ref{fgr:XRD_Charged} show the XRD spectra for \ce{Zn} electrodes which have been discharged to 40\% DoD and subsequently charged in each electrolyte. The XRD results show a clear trend in the precipitation products as a function of electrolyte pH and total chloride content. 

In Figure \ref{fgr:XRD}, the dominant phase in electrolytes E4 and E6 is the layered zinc hydroxide chloride simonkolleite~\cite{Hawthorne2002}, \ce{ZnCl2*4Zn(OH)2*H2O}, with small quantities of \ce{ZnCl2*2NH3} also observed. In electrolyte E7, the total chloride concentration and pH increase and \ce{ZnCl2*2NH3} becomes the majority phase observed. In electrolyte E8, a clear change in the nature of the discharge products in the slightly alkaline environment is observed. Small quantities of both \ce{ZnCl2*2NH3} and \ce{ZnCl2*4Zn(OH)2*H2O} are present, but a change in the primary phase is most obvious. This is evidenced by the evolution of a sharp reflection at approx. 8.35$\degree$ 2$\theta$ (d=10.68\AA) and accompanied by a broad reflection at around 9.7$\degree$ 2$\theta$. These features could not be identified with reference to the available databases. 

\begin{figure*}[t]
  \includegraphics[width=1\textwidth]{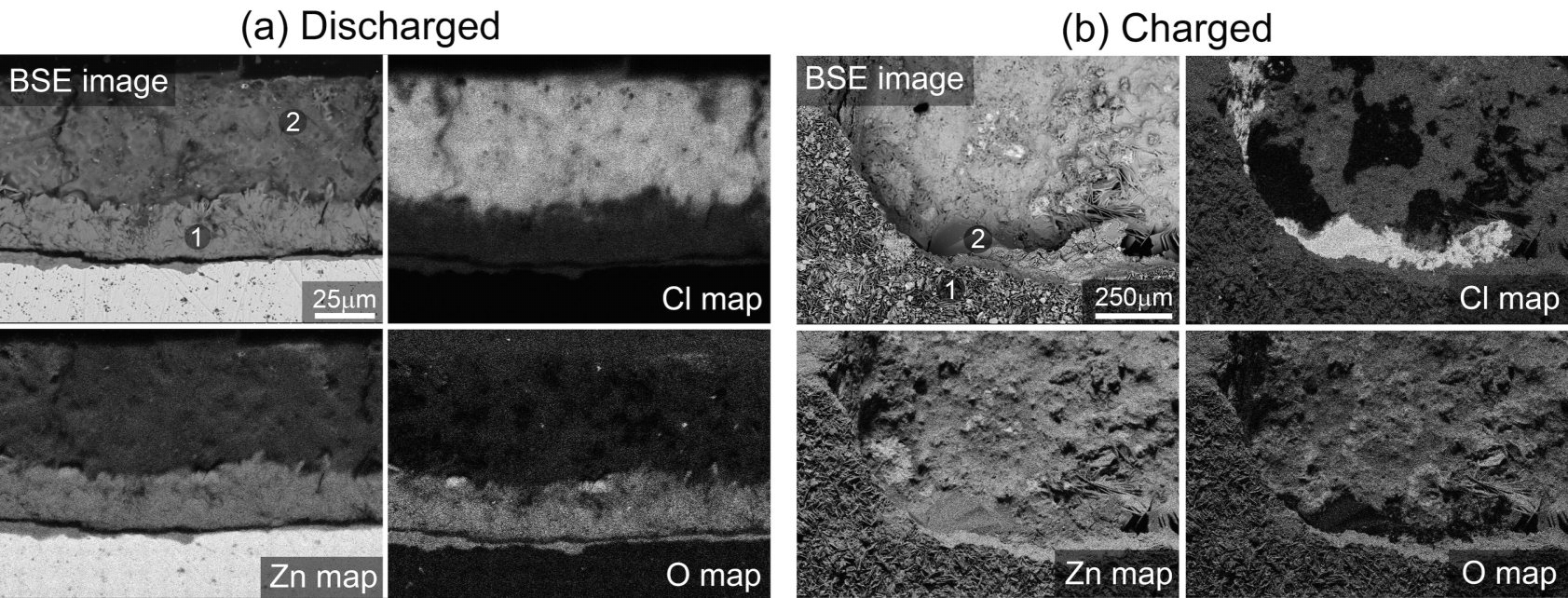}
  \caption{SEM and EDS analysis showing (a) the cross-section of a \ce{Zn} electrode after discharge and (b) the surface of a \ce{Zn} electrode discharged and charged in electrolyte E6. There is a separation of chloride and oxide phases in the discharged electrode. The charged electrode is covered by significant quantities of zinc precipitates. The relative fractions of Zn, Cl, O and N (measured via EDS) for the positions labelled "1" and "2" on each figure are presented in a table in supplementary information$^\dag$.} 
  \label{fgr:SEM_Cycle}
\end{figure*}

It is well-established that layered hydroxide phases, including \ce{ZnCl2*4Zn(OH)2*H2O}~\cite{Arizaga2012}, can intercalate charged and neutral species with concomitant demonstration of an expanded unit cell in the c-direction, corresponding to an expansion of the metal hydroxide layer spacing. Analysis of the E8 diffraction pattern (Figure \ref{fgr:XRD}) yields five sharp diffraction lines which can be fitted as the first five indexes of \{00l\} reflections for a hexagonal unit cell of c=10.68\AA. We note that this accords well with the value of 10.8\AA{} reported by Ar\'{i}zaga \cite{Arizaga2012} for the inter-layer spacing in an ammonia-intercalated sample of \ce{Zn5(OH)8Cl2.H2O}. This interpretation is supported by the observation of nitrogen in this phase by EDS (see supplementary information$^\dag$) and by the thermodynamic model, which predicts that the concentration of \ce{NH3} in the electrolyte increases at higher pH values (Figure \ref{fgr:Speciation}(b)). Assuming the formation of such a pillared hydroxide phase, the broad reflection at around 9.7$\degree$ 2$\theta$ can then be interpreted as arising from incompletely intercalated particles of the same phase. 

One important result of this analysis is that there is no identifiable \ce{ZnO} phase present in any of the discharged samples. Instead, the precipitated phases are dominated by a mixture of zinc hydroxide chlorides. As noted in Table \ref{tbl:OverallReactions}, this alters the overall reaction of the cell and consumes electrolyte as an active component.

Figure \ref{fgr:XRD_Charged} shows the XRD spectra of samples that are first discharged to 40\% DoD and then charged back to their original state-of-charge in each electrolyte. The charged electrodes show the same progression of products from \ce{ZnCl2*4Zn(OH)2*H2O} in E4 and E6, to \ce{ZnCl2*2NH3} in electrolyte E7, and a layered phase in E8. However, the signals from \ce{Zn} metal are strongly reduced. This indicates that the dissolution of the precipitated \ce{Zn} products is suppressed during the charging process, which would have significant consequences for the reversibility of the battery. SEM/EDS characterization of the electrodes give further insight into this observation.

Figure \ref{fgr:SEM_Cycle} shows (a) the cross-section of a \ce{Zn} electrode discharged in electrolyte E6 and (b) the top-down view of the discharged-charged electrode. Both cross-section and top-down views of the SEM/EDS data are presented, so as to give a full insight into the sample microstructure. The relative fractions of Zn, Cl, O and N (measured via EDS) for the positions labelled "1" and "2" on each figure are presented in a table in supplementary information$^\dag$, along with SEM/EDS data for the other samples (E4, E7, and E8).

Figure \ref{fgr:SEM_Cycle}(a) shows that there is a clear separation between layers of chlorine-rich and oxygen-rich phases during discharge. The point EDS composition measurements (supplementary information$^\dag$) further support the observations by XRD that these phases are \ce{ZnCl2*2NH3} and \ce{ZnCl2*4Zn(OH)2*H2O}. The observed phase layering supports the hypothesis that local changes in electrolyte concentration affect the composition of the precipitation product. In this case, the precipitation of a chlorine-rich phase reduces the local concentration of chlorides in the electrolyte, thus favoring the shift towards an oxygen-rich phase. This theory is supported by the thermodynamic analysis in Figures \ref{fgr:Speciation} and \ref{fgr:Thermo2D} and our existing work~\cite{Clark2017}. However, we note that the SEM cross-section indicates the distribution of the phases in space but not in time. Therefore, additional research investigating precipitation at various states of discharge could give further insight into the time-dependent phase formation.

Figure \ref{fgr:SEM_Cycle}(b) shows that after charging, the products that precipitated during the discharge process are not redissolved and deposited as Zn metal, as would be expected for a reversible electrode reaction. Instead, additional material deposits on the Zn electrode which corresponds to the chemical composition of the simonkolleite phase. This is further supported by the cross-sectional image for the recharged cell using electrolyte E4 in the supplementary information$^\dag$. The kinetics of \ce{ZnO} dissolution are known to be sluggish in neutral electrolytes~\cite{Zhang1996}, but this cannot be solely responsible for the limited reversibility of the electrode. The model-based analysis identifies a few mechanisms which can contribute to this observation.

First, there is some inertia in electrolyte behavior between the discharging and charging processes. Because the volume of electrolyte in the cell is comparatively large, it takes time for the electrolyte to transition between quasi-steady-state compositions and precipitation can continue even after charging begins. Second, the same buffering mechanism that works to stabilize the pH during discharge can actually act as a self-braking mechanism in the dissolution of zinc precipitates. When zinc metal is deposited from \ce{Zn(NH3)_{\textit{x}}^{2+}} complexes, ammonia is released into the solution. \ce{NH3} acts as a proton acceptor and stabilizes the pH in the near-neutral regime, as discussed in the section on Equilibrium Thermodynamics. As is evident in Figures \ref{fgr:Speciation} and \ref{fgr:Thermo2D}, maintaining a stable near-neutral pH keeps the electrolyte in a state of low zinc solubility, thereby limiting the dissolution of the precipitated products. Only when either the concentration of dissolved zinc drops to very low levels or the pH becomes more acidic, will the precipitated products dissolve. This mechanism is observed in the cell-level continuum simulations. 

\begin{figure}[t]
  \includegraphics[width=1.0\linewidth]{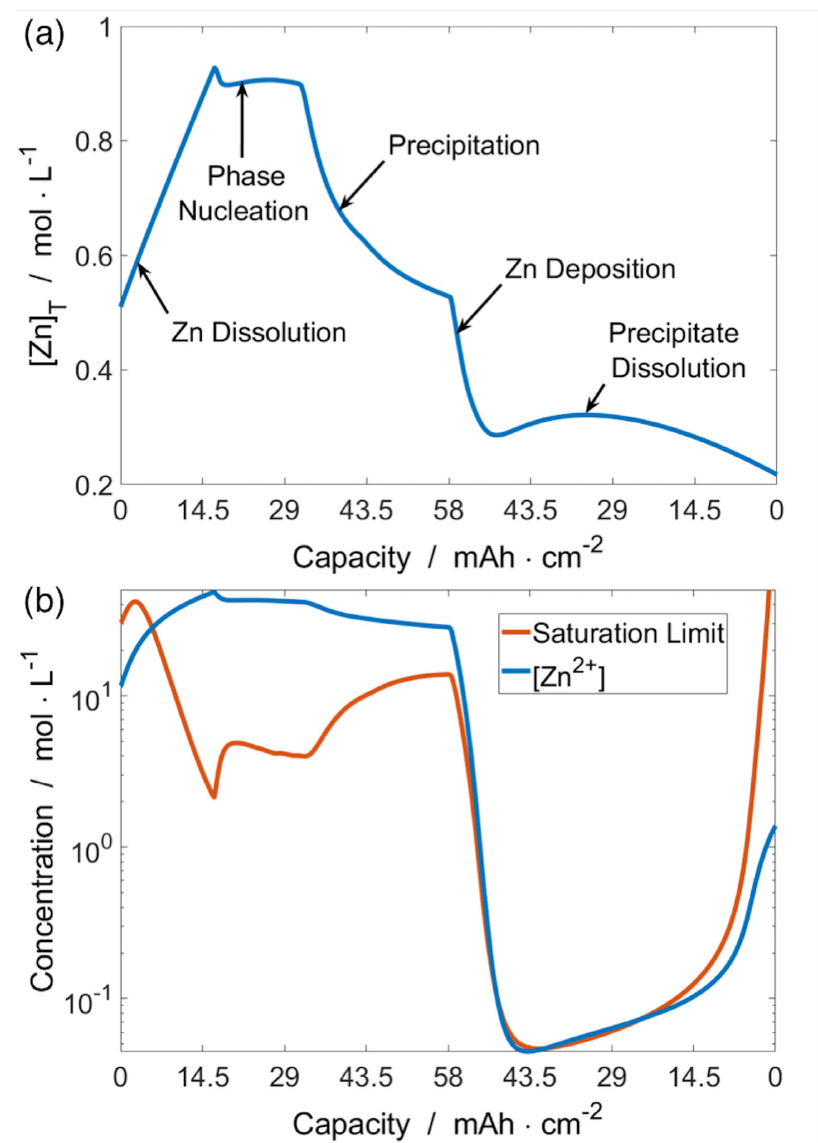}
  \caption{(a) Total zinc concentration in the electrolyte and (b) [\ce{Zn^{2+}}] concentration compared with the saturation limit at the front of the \ce{Zn} electrode over a single discharge-charge cycle. During charging, the simultaneous loss of \ce{Zn^{2+}} (due to deposition) and gain in \ce{NH3} (released from \ce{Zn(NH3)_x} complexes) creates a self-braking effect that slows the re-dissolution of the precipitated zinc products.}
  \label{fgr:Dissolution}
\end{figure}

Figure \ref{fgr:Dissolution} shows (a) the total concentration of zinc in the electrolyte and (b) a comparison of \ce{Zn^{2+}} concentration and the saturation limit over a single simulated galvanostatic cycle in electrolyte E6. In Figure \ref{fgr:Dissolution}(a), the stages of discharge and charge are clearly distinguishable. At the beginning of discharge, the total concentration of zinc in the electrolyte rises as it becomes saturated, levels off as solid zinc phases nucleate, and falls as those phases precipitate. When the cell is charged, the total concentration of zinc falls as it is redeposited on the \ce{Zn} metal electrode and begins to stabilize when the precipitated zinc products start to dissolve. However, the zinc provided by the products dissolution is not enough to fully recover, resulting in a net loss of zinc from the electrolyte.

Figure \ref{fgr:Dissolution}(b) shows the concentration and solubility limit of \ce{Zn^{2+}} at the surface of the Zn electrode. In this figure, the stages of discharge are also clearly distinguishable. At the beginning of discharge, the \ce{Zn^{2+}} concentration increases as the \ce{Zn} electrode dissolves. When the solution becomes supersaturated with zinc, solid phases nucleate and the \ce{Zn^{2+}} concentration falls as they precipitate. When the cell is charged, the concentration of zinc in the electrolyte drops further as the \ce{Zn} electrode is plated. Although the concentration of \ce{Zn^{2+}} also decreases as it is deposited, the amount of free \ce{NH3} rises as it is released from \ce{Zn(NH3)_{\textit{x}}^{2+}} complexes. The combination of a low concentration of \ce{Zn^{2+}} and an excess of \ce{NH3} conspire to create a self-braking effect and slow the dissolution of the zinc precipitates. Only when the acidic front from the air electrode reaches the \ce{Zn} electrode at the end of charging does the solubility increase.

In summary, the ex-situ characterization of the \ce{Zn} electrodes shows that \ce{ZnO} is not the dominant discharge product in the investigated L-ZAB cells. Rather, the zinc precipitate phases are dominated by a mix of zinc hydroxide chlorides. Furthermore, the reversibility of the precipitated products appears to be limited. When the cell is charged, the zinc products are slow to dissolve, and a significant quantity of precipitated material remains on the electrode at the end of charging. These observations are in accord the theory of the system, and the mechanisms driving these processes are elucidated by the cell models.

\section{Conclusions}

Near-neutral \ce{ZnCl2-NH4Cl} electrolytes could extend the lifetime of rechargeable ZABs by minimizing the effects of electrolyte carbonation, but these electrolytes bring new challenges that must be addressed in material development and cell design. Model-based analysis makes two important predictions about L-ZAB operation: the pH can become strongly acidic during charging and the dominant precipitation product is not \ce{ZnO}. Both of these predictions are experimentally observed.

Operando pH measurements obtained during cell cycling confirm that the electrolyte can become strongly acidic. According to our model, this effect is driven by the slow diffusion and low concentration of \ce{NH3} in the electrolyte. Acidic electrolyte environments can accelerate catalyst degradation and material corrosion, thereby limiting the lifetime of the battery.

The precipitation of zinc products is problematic for L-ZAB design and operation. Ex-situ XRD, SEM, and EDS measurements confirm the model-based prediction that the dominant solid discharge product is not \ce{ZnO} but a pH-dependent mix of zinc hydroxide chloride phases. Furthermore, although the pH buffer is needed to stabilize the performance of the air electrode, it slows the dissolution of zinc precipitates during charging and limits the reversibility of the battery.

As a topic for future research, forced convection of the electrolyte could help address some of these challenges. A flow cell configuration would reduce the mass transport limitations in the buffer solution and limit the precipitation of problematic zinc salts, but the energy density of the system would be significantly reduced. Nonetheless, well-designed LeClanch\'{e} zinc-air flow batteries could potentially find use in stationary energy storage systems. 


\section*{Conflicts of Interest}

The authors declare no conflicts of interest.

\section*{Acknowledgement}

This work has received funding from the European Union's Horizon 2020 research and innovation program under grant agreement No. 646186 (ZAS! project) and from the Basque Country Government (ELKARTEK 2017 program). The support of the bwHPC initiative through the use of the JUSTUS HPC facility at Ulm University is acknowledged. The electrochemical characterization in this work has been done in the frame of the Doctoral Degree Program in Chemistry by the Universitat Aut\`onoma de Barcelona.




\bibliography{Mendeley} 
\bibliographystyle{rsc}
\end{document}